\documentclass{vospwks2007}

\title{Results of an analysis of SDSS galaxies in the VO}
\author[1]{W. Schoenell}
\author[2,3]{M. Cervi\~no}
\author[1]{R. Cid Fernandes}
\author[4]{A. Mateus}
\author[5]{E. Terlevich}
\author[5]{R. Terlevich}
\author[5]{F. de los Santos}
\author[1,5]{J.P. Torres-Papaqui}
\author[2,3]{V. Luridiana}
\affil[1]{Universidade Federal de Santa Catarina (UFSC; Brazil)}
\affil[2]{Spanish Virtual Observatory (SVO; Spain)}
\affil[3]{Instituto de Astrof\'\i sica de Andaluc\'\i a (IAA-CSIC; Spain)}
\affil[4]{Instituto de Astronomia, Geof\'\i sica e Ci\^encias Atmosf\'ericas (IAG-USP; Brazil)}
\affil[5]{Instituto Nacional de Astrof\'\i sica, \'Optica y Electr\'onica (INAOE; Mexico)}

\begin{document}


\maketitle

\begin{abstract}
We present here the VO access to the results of an analysis of the spectra of Sloan Digital Sky Survey (SDSS) galaxies performed with 
the STARLIGHT code by \cite{Cidetal05}. The results include for each galaxy the original SDSS spectrum, the best-fit synthetic 
spectrum, the star formation history, the pure emission line spectrum corrected for the underlying stellar population (in SDSS emission 
line galaxies) and the intensity of several emission/absorption lines. The database will be accessible from the PGos3 server and it 
will be released at the end of summer 2007.
\end{abstract}

\section{Introduction}

The interoperability concept promoted by the Virtual Observatory (VO) is not only limited to data sharing, but it also includes 
enhancing the possibility to analyze these data (usually by means of comparison with theoretical models) and to infer relevant physical 
quantities. This task can only be performed if front-end VO interfaces (mostly, applications) are able to provide access to theoretical 
data (see Rodrigo's et al. contribution in these proceedings) and if analysis tools are implemented in a VO compliant fashion. 
Theoretical and observational data must not be hardwired in the VO-analysis tools: rather, they should be obtained from VO services. 
This is the only way to minimize the model-dependent bias in the analyzed data.

The use of VO services to input theoretical and observational data into VO analysis tools creates diversity and improves productivity: 
since the VO provides a homogeneous structure for data sharing (VOTables), it allows to design automatic tools and workflows that can 
use different sets of observational and theoretical data without making program drivers for each specific data set. Once data providers 
distribute their results in VOTables, they can use these analysis tools for further research.

Although this is in principle the most desirable scenario, there are several difficulties that make this kind of workflow difficult to 
establish, mainly: (i) There is no user-friendly way to recover the credits of the data used in current VO applications; an issue 
mentioned several times during this meeting and especially relevant for theoretical studies. (ii) The access and exploration of 
VO-enabled services with theoretical datasets\footnote{Examples are the PEGASE-HR (see Ph. Prugniel's contribution in these 
proceedings) and PGos3 databases (see Cervi\~no's et al. contribution in these proceedings) for evolutionary synthesis models at 
{\tt{http://vo.obspm.fr/cgi-bin/siap/pegasehr.pl}} and {\tt{http://ov.inaoep.mx/pgos3}} respectively.} are not implemented in most 
VO-applications, with VOSpec and VOSED as the only exceptions.

In this contribution we describe the first stages of the implementation of the analysis tool STARLIGHT \citep{Cidetal05} in the VO 
framework.

\section{The STARLIGHT code}

The STARLIGHT code is designed to obtain the best fit to an observed spectrum, $O_{\lambda}$, taking into account the corresponding 
error $\sigma_{\mathrm{obs}}$, with a theoretical model spectrum, $M_{\lambda}$, using a Markov Chains Monte-Carlo algorithm with 
simulated annealing. The code finds the minimum $\chi^2$,

\begin{equation}
\chi^2 = \sum_{\lambda} \left(\frac{O_{\lambda} - M_{\lambda}}{\sigma_{\mathrm{obs}}}\right)^2,
\end{equation}

\noindent and obtains the corresponding physical parameters of the modeled spectrum: (i) the star formation history, $x_{j}$, as a 
function of a base of $N_{\mathrm{SSP}}$ Single Stellar Population (SSP) models normalized at $\lambda_{0}$, $b_{j,\lambda}$ (ii) the 
extinction coefficient of predefined extinction laws, $r_{\lambda}$, and (iii) the velocity dispersion $\sigma_{v}$, which obey the 
relation:

\begin{equation}
M_{\lambda} = M_{\lambda_{0}} \left(\sum_{j=1}^{N_{\mathrm{SSP}}} x_{j}\, b_{j,\lambda}\, r_{\lambda}\right)\, \otimes\,G(v_{*},\sigma_{*}).
\end{equation}

With this technique, we have obtained the physical parameters of the galaxies of the SDSS data release 2 
\citep{sloan1,sloan2,sloan3,sloan4}, using the library of SSP models by \cite{BC03} as database. Since the modeled spectrum provides 
the stellar continuum of the galaxy, we can also measure emission line intensities and equivalent widths. This huge computational 
effort resulted in a unique database with more than  half a million galaxies, which occupies close to 500 GB of data.

\section{VO implementation}

The VO implementation of STARLIGHT allows to explore different issues in the VO, from describing the results of the data analysis 
(including references to input data, models and processes) to establishing a VO workflow using theoretical VO services.

We created a preliminary web service to access and manipulate this database and obtain the corresponding VOTables at {\tt 
http://www.starlight.ufsc.br/}. 
Users can perform queries on data by SQL commands or spatial selections. The final VO service is expected to be available at the 
servers {\tt http://www.starlight.ufsc.br/} and {\tt http://ov.inaoep.mx}.

Currently we are working on two different aspects:

\begin{enumerate}
\item To produce tools to explore the database in a VO compliant way instead of a Web service. First essays using TSAP (Theoretical 
Spectral Access Protocol, see Rodrigo's et al. talk in these proceeding for extensions to non spectroscopic data) are not satisfactory 
enough due to the database complexity. At this point, a more general, recursive, protocol is needed.
\item To allow the STARLIGHT code to obtain SSP databases from (non-local) VO services. We are making this implementation using SSP 
models included in PGos3 database (see Cervi\~no's et al. contribution in these proceedings) as reference.
\end{enumerate}

However, other issues remain to be addressed. Maybe the most important ones are:
\begin{enumerate}
\item How to manage references to VOTables nested in VOTables: the description of the star formation history of a galaxy needs nested 
references to VOTables with both original data and the SSP database used. At the same time, the SSP database used would need references 
to the assumed isochrones and atmosphere models, and so on.
\item How to know the coverage, uncertainties and range of validity of different SSP models in an automatic way. This issue can be 
partially addressed by the use of the current proposed IVOA recommendations on data model characterization and the spectral data model 
(see Cervi\~no \& Luridiana contribution in these proceedings for their possible applications to SSP models).
\end{enumerate}

\section*{Acknowledgments}

We acknowledge the INAOE and LAEFF for financial support at different stages of this project. We acknowledge Carlos Rodrigo for useful 
comments on the manuscript. We also acknowledge Miguel Mart\'\i nez and the Computer Services of INAOE for infrastructure management 
and technical support during the elaboration of this project.  The STARLIGHT project is supported by the Brazilian agencies CNPq, CAPES 
and FAPESP. This research has made use of the Spanish Virtual Observatory supported from the Spanish MCyT through grants AyA2005-04286, 
AyA2005-24102-E. It has also been supported by the Spanish MCyT through the project AYA2004-02703. MC is supported by a {\it Ram\'on y 
Cajal} fellowship.
Funding for the Sloan Digital Sky Survey has been provided by the Alfred P. Sloan Foundation, the Participating Institutions, the 
National Aeronautics and Space Administration, the National Science Foundation, the U.S. Department of Energy, the Japanese 
Monbukagakusho, and the Max Planck Society.

\end{document}